\documentstyle[12pt]{article}
\newcommand{\extraspace}{\addtolength{\abovedisplayskip}{2mm}
                        \addtolength{\belowdisplayskip}{2mm}
                        \addtolength{\abovedisplayshortskip}{2mm}
                        \addtolength{\belowdisplayshortskip}{2mm}}
\newcommand{\be}{\begin{equation}\extraspace}

\newcommand{\ee}{\end{equation}}

\newcommand{\bea}{\begin{eqnarray}\extraspace}
\newcommand{\beastar}{\begin{eqnarray*}\extraspace}
\newcommand{\eea}{\end{eqnarray}}
\newcommand{\eeastar}{\end{eqnarray*}}

\newcommand{\nonu}{\nonumber \\[2mm]}

\newcommand{\str}{\rule[-2.5mm]{0mm}{7mm}}
\newcommand{\Str}{\rule[-3.5mm]{0mm}{8mm}}

\newcommand{\half}{{\textstyle \frac{1}{2}}}
\newcommand{\quart}{\frac{1}{4}}

\newcommand{\si}{\sigma}

\newcommand{\de}{\delta}
\newcommand{\cd}{c^{\dagger}}
\newcommand{\Sd}{S^{\dagger}}

\newcommand{\up}{\uparrow}
\newcommand{\down}{\downarrow}

\newcommand{\la}{\lambda}
\newcommand{\lae}{\lambda^{(1)}}
\newcommand{\lat}{\lambda^{(2)}}

\newcommand{\np}{Nucl.Phys.\ }
\newcommand{\prl}{Phys.Rev.Lett.\ }

\newcommand{\cmp}{Comm.Math.Phys.\ }

\begin{document}


\baselineskip=12pt

\hfill {PUPT-1420}

\vskip 1mm

\hfill {CERN-TH.6982/93}
\vskip 1.0cm
\begin{center}

{\LARGE Complete Solution of a Supersymmetric}\\
\vspace{5mm}
{\LARGE Extended Hubbard Model}

\vskip 1.0cm

{\large Kareljan Schoutens}

\vskip .5cm

\baselineskip=15pt

{\sl Joseph Henry Laboratories, Princeton University \\
     Princeton, NJ 08544, U.S.A.}

\vskip .3cm
{\sl and}
\vskip .3cm

{\sl TH-Division, CERN \\
     CH-1211 Geneva 23, SWITZERLAND}
\vskip 2.0cm

{\bf Abstract}

\end{center}

\vspace{.2cm}

\noindent
\baselineskip=17pt
We show that the recently constructed exact solution of the
$SU(2|2)$ extended Hubbard model on a one-dimensional lattice
provides a complete set of $4^L$ eigenstates of the hamiltonian,
where $L$ is the length of the lattice.

\vfill
\noindent PUPT-1420  \hfill

\noindent CERN-TH.6982/93  \hfill

\noindent September 1993


\newpage

\baselineskip=17pt

\section{Introduction}
\setcounter{section}{1}

The dynamics of a system of strongly correlated electrons
on a lattice is most directly described by a hamiltonian
that acts on the microscopic quantum states of the system.

The simplest example is the hamiltonian of the Heisenberg
$XXX$ magnet, which describes spin-spin interactions of
electrons that are nearest neighbors on the lattice. In
this model the electrons are not mobile; one assumes that
each lattice site has either a spin-up or a spin-down
electron, so that the total number of independent states,
for a lattice of $L$ sites, is equal to $2^L$.

If one is interested in modelling phenomena such as
(high-$T_c$) superconductivity, one allows empty or doubly
occupied sites and includes in the hamiltonian
kinetic (or hopping) terms which allow the electrons to move.
A simple example is the so-called $t$-$J$ hamiltonian,
which acts in a space of states (of dimension $3^L$)
in which doubly occupied sites are forbidden.  Besides the kinetic
terms, the $t$-$J$ hamiltonian contains nearest neighbor
density-density and spin-spin couplings.

A further refinement can be made by allowing both empty and
doubly occupied sites (enlarging the basis of the Hilbert space
to $4^L$ states). The probability of finding
doubly occcupied sites can be adjusted through an on-site
potential term, the so-called Hubbard potential, which
contributes an amount $U$ of potential energy for each
doubly occupied site. In the so-called Hubbard model there
are no interactions besides this potential term, but it is
very natural%
\footnote{See for example Hubbard's original paper \cite{hub},
          where estimates for the strengths of some additional
          interaction terms are given.}
to include nearest neighbor interactions such
as the density-density ($V$) and spin-spin ($J$) couplings
of the $t$-$J$ model and additional interactions such as a
bond-charge term ($X$) and a pair-hopping term ($Y$).
One certainly expects
that a hamiltonian that has each of these couplings tuned to
a realistic strength will give an accurate description of the
physics of the electron gas.

In formula, this general hamiltonian takes the following form
\bea
  H &=&  -t \sum_{<ij>,\si}
         (\cd_{i,\si} c_{j,\si} + \cd_{j,\si}c_{i,\si})
\nonu &&
         + V \sum_{<ij>} (n_i-1)(n_j-1)
         + J \sum_{<ij>} (S_i^z S_j^z + \half \Sd_i S_j
         + \half \Sd_j S_i)
\nonu &&
         + X \sum_{<ij>,\si}
           (\cd_{i,\si} c_{j,\si} + \cd_{j,\si}c_{i,\si})
           (n_{i,-\si}+n_{j,-\si})
\nonu &&
         + Y \sum_{<ij>}
           (\cd_{i,1}\cd_{i,-1}c_{j,-1}c_{j,1} +
           \cd_{j,1}\cd_{j,-1}c_{i,-1}c_{i,1})
\nonu &&
         + U \sum_i (n_{i,1}-\half)(n_{i,-1}-\half) \ ,
\label{ham}
\eea
where $<\! ij\! >$ denotes the nearest neighbor relation on the
lattice. The operators $\cd_{i,\si}/c_{i,\si}$ ($i=1,2,\ldots,L$,
$\si= \pm 1$) create/annihilate electron states. They satisfy
$\{ \cd_{i,\si},c_{j,\tau} \} = \delta_{i,j} \delta_{\si,\tau}$.
The space of states is a Fock space with the pseudo vacuum
$\mid \! 0 \rangle$ defined by $c_{i,\si} \! \mid \! 0 \rangle =0$.
The operator $n_{i,\si}=
c^\dagger_{i,\si} c_{i,\si}$ is the number operator for electrons
with spin $\si$ on site $i$ and $n_i = n_{i,1} + n_{i,-1}$.
The $SU(2)$ spin operators are given by
$S^z_i = \half (n_{i,1} - n_{i,-1})$, $\Sd_i = \cd_{i,-1}c_{i,1}$
and $S_i = \cd_{i,1} c_{i,-1}$.

The general hamiltonian (\ref{ham}) is obviously hard to
analyze. However, life becomes simpler if one assumes
certain relations between the coefficients of the interaction
terms. In general, exact results are most
easily obtained in models that have a high degree of symmetry,
and one may therefore look for special cases where the hamiltonian
(\ref{ham}) admits symmetries that go beyond the $SU(2)$
spin rotations. Following this line of thought,
we observe that for each choice of the kinematical
constraint on the electron states (allowing 2,3 or 4 states
per site), there exists a hamiltonian that is maximally
symmetric, {\it i.e.}\ that is invariant under all global,
unitary rotations among the elementary electron states.
The Lie (super-)algebras that describe these symmetries are
$SU(2)$, $SU(2|1)$ and $SU(2|2)$, respectively, where
the grading expresses the fact that the single electron states
have fermionic statistics.

The first two maximally symmetric models are well-known: the
$SU(2)$ symmetric case is the $XXX$ Heisenberg
magnet and the model with $SU(2|1)$ symmetry is the $t$-$J$ model
at the supersymmetric point $J=2t$. The model with $SU(2|2)$
symmetry, which is the logical third in this list, has been
considered only recently \cite{eks1}. Its hamiltonian is of the form
(\ref{ham}) where $V=-t/2$, $J=2t$, $X=t$, $Y=-t$, and $U=-dt$,
where $d$ is the number of nearest neighbors of each lattice
site. The physical properties of this model are quite remarkable:
it was shown in \cite{eks2} that the groundstate in 1,2 or 3
dimensions is superconducting, even when a moderately repulsive
Hubbard potential term is added to the hamiltonian.

The superconductivity in the $SU(2|2)$ model comes about through
the Bose-Einstein (BE) condensation of local electron pairs
(bipolarons), which is in contrast to the usual BCS mechanism.
It has been observed (see for example \cite{flr,ran}) that the
properties of `BE superconductors' agree rather nicely with the
observed phenomenology of the high-$T_c$ superconductors (this is
for both the superconducting and the normal state). We therefore
expect the $SU(2|2)$ model, which can be viewed as an `ideal BE
superconductor'%
     \footnote{`Ideal' in the sense that {\it momentum-zero} electron
         pairs only interact via a chemical potential term.
         For non-zero momentum pairs this is no longer the case.},
can play an important role in the quest for a
theory for high-$T_c$ superconductivity. We refer to a forthcoming
paper \cite{eks4} for further discussion of the physics of the model.

{}From a mathematical point of view the three maximally symmetric
electron models are relatively easy to work with: in particular,
they share the property of being exactly solvable in 1 dimension.
(The Hubbard model, which is not maximally symmetric, also
has this property \cite{liebwu}.) The exact solution of the $SU(2|2)$
model employs a three-fold nested, graded Bethe Ansatz. It was
presented in \cite{eks3}, see also \cite{kul}, and has been
further worked out in \cite{ek}.

The Bethe Ansatz being an {\it Ansatz}, there is
of course the issue whether or not it leads to a {\it complete}
solution of a given model. A more precise question is whether
the Bethe Ansatz, {\it together with the global symmetries of
the model}, provides a complete set of eigenstates of the
hamiltonian (complete in the sense that these states form
a basis for the Hilbert space of the model).
For the $XXX$ model this issue was already discussed in
Bethe's original paper \cite{bethe}. For the Hubbard
model and for the supersymmetric $t$-$J$ model completeness
has been established only recently \cite{eks0,karfoe}.
In this paper we shall extend these results to the $SU(2|2)$
extended model and explicitly discuss the counting of the
eigenstates.

The paper is organized as follows. In the next section we
briefly review the Bethe Ansatz and we discuss the so-called
string solutions to the Bethe equations. We derive a formula
for the number of solutions that can be constructed by combining
a specific set of string solutions, and we specify the
dimensions of the $SU(2|2)$ multiplets that are present
in the spectrum. The counting of the eigenstates is discussed
in detail in section 3.

\section{Analyzing the Bethe equations}
\setcounter{equation}{0}

The exact solution of the $SU(2|2)$ extended Hubbard model
on a one-dimensional lattice starts from the observation
\cite{eks1} that the hamiltonian $H_0$ (which is as in (\ref{ham})
with $t=1$, $V=-1/2$, $J=2$, $X=1$, $Y=-1$ and $U=-2$)
acts as minus the sum of graded permutations $\Pi^g_{ij}$
of the electronic states on neighboring sites $i$, $j$ :
\be
H_0 = - \sum_{<ij>} \Pi^g_{ij} \ .
\label{hperm}
\ee
Lattice models for which the hamiltonian acts as a
(graded) permutation were first considered in \cite{sut}.
We shall here use the so-called FFBB representation
of the Bethe Ansatz solution, which was presented
in detail in \cite{eks3} (see also \cite{kul}). The
details of the Bethe Ansatz approach are rather
complicated, but the final result is easily explained
as follows. The (nested) Bethe Ansatz provides a
multi-parameter family of states, written as
\be
| \, \la_1,\ldots,\la_{N_l+N_e}\, ; \, \lae_1, \ldots
  \lae_{N_e} \, ; \, \lat_1, \ldots, \lat_{N_d} \, \rangle \ ,
\label{es}
\ee
which turn into exact eigenstates of the hamiltonian $H_0$
when the spectral parameters $\{ \la_k, \lae_j, \lat_m\}$
solve the following set of algebraic equations
\bea
&& \left( {\la_k -{i \over 2} \over \la_k + {i \over 2}} \right)^L =
\prod_{j=1}^{N_e}
{\la_k - \lae_j + {i \over 2} \over \la_k - \lae_j - {i \over 2}}
\; \prod_{l=1, l\neq k}^{N_l+N_e}
{\la_l - \la_k + i \over \la_l - \la_k - i} \ ,
\qquad k=1,\ldots,N_l+N_e
\nonu
&& \prod_{k=1}^{N_l+N_e}
{\la_k - \lae_j + {i \over 2} \over \la_k - \lae_j - {i \over 2}} =
\prod_{m=1}^{N_d}
{\lat_m - \lae_j + {i \over 2} \over \lat_m - \lae_j - {i \over 2}} \ ,
\qquad j=1,\ldots,N_e
\nonu
&& \prod_{l=1, l\neq m}^{N_d}
{\lat_l - \lat_m + i \over \lat_l - \lat_m - i}  =
\prod_{j=1}^{N_e}
{\lat_m - \lae_j - {i \over 2} \over \lat_m - \lae_j + {i \over 2}} \ ,
\qquad m=1,\ldots,N_d  \ .
\label{bethe}
\eea
These equations are called the {\it Bethe equations}\ for this
specific Bethe Ansatz.
By the symbols $N_l$, $N_e$, and $N_d$ we denote the number
of local electron pairs, of single electrons and of
single spin-down electrons, respectively. The total
numbers of spin-up ($N_\up$) and spin-down ($N_\down$)
electrons are thus given by
\be
N_\up = N_l + N_e - N_d \ , \qquad
N_\down = N_l + N_d \ .
\ee
The energy of the eigenstate (\ref{es})
for a given solution $\{\la_k,\lae_j,\lat_m\}$ of the Bethe
equations is given by
\be
E = \sum_{k=1}^{N_l+N_e} {1 \over \la_k^2+\quart} - L \ .
\ee

Before we come to the description of the solutions to the
Bethe equations, we wish to make two important remarks,
which are both relevant for the counting of eigenstates
that we are going to perform.

The first remark concerns an especially nice property of
the Bethe Ansatz eigenstates, which is that they are lowest
weight states of the full symmetry algebra $SU(2|2)$
\cite{eks3}. This implies that each Bethe Ansatz eigenstate
comes with an entire multiplet of eigenstates of the hamiltonian.
The dimensions of these multiplets, which are derived by using
the representation theory of the algebra $SU(2|2)$, are given
below in the equation (\ref{dims}).

The second remark concerns the norm of the Bethe Ansatz eigenstate
for a given solution $\{ \la_k,\lae_j,\lat_m\}$. For the simpler
case of the non-nested bosonic Bethe Ansatz there exists an elegant
formula for the norm of all Bethe Ansatz eigenstates
\cite{vlad}, which in particular implies that this norm is
non-vanishing for so-called regular solutions to the
Bethe equations. However, for the present situation a closed
expression for the norm of the Bethe Ansatz eigenstates is
not known, and we have to rely on other arguments to find out
which solutions to the Bethe equations (\ref{bethe}) give rise
to non-vanishing eigenstates. We propose the term {\it spurious
solution}, for a solution to the Bethe equations that
corresponds to a vanishing eigenstate.

Let us now focus on the solutions of the equations (\ref{bethe}).
We will interpret general solutions of (\ref{bethe}) as combinations
of so-called {\it string configurations}, which are configurations
in the complex $\la$-plane that are close to the {\it string solutions}
that solve the Bethe equations in the limit where $L$ tends to infinity.
For the present model, the relevant string solutions are \cite{ek}
\bea
&& \hbox{$\la$-strings} \, : \, \la^{mj}_{\alpha}
   = \la^m_{\alpha} + {i \over 2}(m+1-2j)\, ,
   \quad j=1,\ldots,m
\nonu
&& \hbox{real}\; \lae \, : \, \lae_{\beta}
\nonu
&& \hbox{$\lae$-$\lat$ strings} \, : \, \lat_\gamma \, , \quad
   \la^{(1)j}_\gamma
   = \lat_\gamma + {i \over 2}(3-2j) \ , \quad j=1,2
\nonu
&& \hbox{$\lat$-strings} \, : \, \la^{(2)mj}_{\nu}
   = \la^{(2)m}_{\nu} + {i \over 2}(m+1-2j)\, , \quad j=1,\ldots,m \ ,
\label{str}
\eea
where the {\it string centers}\ $\la^m_\alpha$, $\lae_\beta$,
$\lat_\gamma$ and $\la^{(2)m}_\nu$ are all real numbers. Notice
that the $\lae$-$\lat$ strings each consist of one real rapidity
$\lat_\gamma$ and two complex rapidities $\la^{(1)j}_\gamma$,
$j=1,2$.

We should now determine in what ways these string configurations
can be combined into true solutions of the equations (\ref{bethe}).
This can be done by substituting the string solutions (\ref{str})
into the Bethe equations and extracting a set of equations
for the string-centers. We shall focus on a solution that
contains $M_m$ $\la$-strings of length $m$, $N_1$ real
$\lae$, $N_{12}$ $\lae$-$\lat$ strings and $K_m$ $\lat$-strings
of length $m$.

Imposing periodic boundary conditions implies that the string
centers have to be chosen from a discrete set. We will make the
(standard) `assumption' (see below for the quotation marks) that
this set is parametrized by a set of non-repeating strictly
positive integers, one for each string configuration that is
present in the solution. We will denote the integers corresponding
to the string centers $\la^m_\alpha$, $\lae_\beta$, $\lat_\gamma$,
and $\la^{(2)m}_\nu$ by $I_\alpha^m$, $J_\beta$, $K_\gamma$,
and $N^m_\nu$, respectively. The allowed ranges of these integers
were studied in \cite{ek}. Following this work (with one
modification, see below), we claim that the non-spurious solutions
to the Bethe equations are parametrized by non-repeating integers
satisfying
\bea &&
0 < I^m_\alpha \leq
  L- \sum_{n=1}^\infty (t_{mn} M_n) + N_1 + N_{12}(2-\delta_{m1})
\nonu &&
0 < J_\beta \leq
\sum_{m=1}^\infty \left( M_m - K_m \right)  -N_{12} -1
\nonu &&
0 < K_\gamma \leq
M_1 + 2 \sum_{m=2}^\infty M_m  - N_1 - N_{12} - 2
\nonu &&
0 < N^m_\nu \leq
N_1 - \sum_{m=1}^\infty (t_{mn} K_n) \ ,
\label{ineq}
\eea
where $t_{mn} = 2\, {\rm Min} (m,n) - \de_{mn}$. This
result implies that the total number of solutions with
string multiplicities $\{M_m,N_1,N_{12},K_m\}$ is given by
\bea
&& \hskip -2pt n(\{ M_m \}, N_1, N_{12},  \{ K_m \})=
\prod_{m=1}^{\infty} \left( \begin{array}{c}
   \str L - \sum_{n=1}^\infty (t_{mn} M_n)
    + N_1 + N_{12}(2-\delta_{m1}) \\ \str M_m \end{array} \right)
\nonu && \hspace{1.5cm} \times
\left( \begin{array}{c}
   \str \sum_{n=1}^\infty (M_n - K_n) - N_{12} - 1
   \\ \str N_1 \end{array} \right)
\nonu && \hspace{1.5cm} \times
\left( \begin{array}{c}
   \str M_1 + 2 \sum_{n=2}^\infty M_n - N_1 - N_{12} - 2
   \\ \str N_{12} \end{array} \right)
\nonu && \hspace{1.5cm} \times
\prod_{m=1}^\infty \left( \begin{array}{c}
   \str N_1 - \sum_{n=1}^\infty (t_{mn} K_n)
   \\ \str K_m \end{array} \right) \ .
\label{number}
\eea
Before we continue, several comments on the validity of this
formula are in order.

A first comment concerns the `assumption' that the solutions to the
Bethe equations (\ref{bethe}) are in 1-1 correspondence with sets of
non-repeating integers that satisfy the constraints (\ref{ineq}).
This assumption is similar to
an assumption made in Bethe's original work on the Heisenberg
$XXX$ chain \cite{bethe}, which was followed in later work by many.
However, Bethe immediately pointed out that this assumption gives a
slightly idealized version of the situation, since the actual string
content of solutions can be different from the one implied by this
counting. This phenomenon was carefully studied
in ref. \cite{eks01}, where it was argued that the assumption
is correct up to a redistribution of solutions of one type
into another. (For example: it was found that for increasing
lattice length certain complex 2-magnon string solutions in
the $XXX$ model turn into pairs of real solutions.)
With this in mind, we will continue our discussion on
the basis of the equation (\ref{ineq}).

A second comment concerns the upper limits on the integers
$I^m_{\alpha}, J_{\beta}, K_{\gamma}$ and $N^m_{\nu}$.
Our claim is that the ranges specified in (\ref{ineq}) are
such that they precisely select solutions to the Bethe equations
that correspond to non-vanishing Bethe Ansatz eigenstates.
We have tested this claim by explicitly working out the
$SU(2|2)$ quantum numbers of all lowest weight states for the
model on $L=2,3,\ldots,9$ sites. The multiplicities obtained
are precisely reproduced by the expression (\ref{number}).
There can thus be little doubt that the proposed expression is
correct (the check at $L=9$ involves a total of 262,144 states,
2,578 of which are lowest weight states).

If one is interested in obtaining solutions to the
Bethe equations (without insisting that the corresponding
eigenstates be non-vanishing), one finds a slightly larger
domain for the allowed integers $I^m_{\alpha}, J_{\beta},
K_{\gamma}$ and $N^m_{\nu}$. This observation explains the
difference between the inequalities (\ref{ineq}) and those
reported in \cite{ek}. The following set of rapidities
provides an example of a spurious solution to the Bethe
equations (\ref{bethe}) for $L=3$
\be
\la_1 = \frac{1}{6} \sqrt{3}, \quad
\la_2 = -\frac{1}{6} \sqrt{3}, \quad
\la^{(1)}_1 = \frac{i}{3}\sqrt{3},  \quad
\la^{(1)}_2 = -\frac{i}{3}\sqrt{3}, \quad
\la^{(2)}=0 \ .
\label{sps}
\ee
It can be viewed as a combination of two real $\la$'s and
one $\la^{(1)}$-$\la^{(2)}$ string, with $M_1=2$, $N_{12}=1$.
It is a perfectly sound solution to the Bethe equations
and as such its existence was predicted by the analysis of
\cite{ek}. However, it can easily be seen that this solution
does not correspond to a non-vanishing eigenstate of the
hamiltonian. The reasoning is as follows. We know that the
state has quantum numbers $N_l=0$, $N_\up=1$, $N_\down=1$,
and that its energy is $E=3$. It is easily checked that the
spectrum of the $L=3$ model has only one state with these
characteristics, and that this state is not a lowest weight
state for $SU(2|2)$. Since the theorem of \cite{eks3} guarantees
that all Bethe Ansatz states are lowest weight states, we have
to conclude that the state with spectral parameters (\ref{sps})
is actually vanishing. This result has been confirmed by explicit
evaluation of the Bethe Ansatz eigenstate \cite{fabian}.
Recently, the solution (\ref{sps}) has been generalized to a
large class of spurious solutions \cite{fabian}.

Before closing this section, we come back to the multiplet
structure of the space of eigenstates of the hamiltonian
(\ref{hperm}). We already mentioned that the lowest weight theorem
for Bethe Ansatz eigenstates implies that every Bethe Ansatz
eigenstate gives rise to a multiplet of eigenstates which form a
representation of $SU(2|2)$.
The dimensions of these multiplets can be obtained from the
representation theory of the superalgebra $SU(2|2)$, as explained
for example in \cite{susyrep}. It turns out that the dimensions
can be expressed in terms of the numbers $N_\up$ and $N_\down$.
Depending on the value of $N_\down$, the representations can be
{\it typical} or {\it atypical}. The precise result is
\bea
\hbox{\it typical:} & & \hbox{\rm dim}_{N_\up,N_\down}
    = 16 (L-N_\up-N_\down+1) (N_\up-N_\down+1)
\nonu
& &  \qquad \qquad \qquad \qquad \hbox{\rm if}\ \
     N_\up\geq N_\down \geq 2, \ N_\up+N_\down \leq L\ ,
\nonu
\hbox{\it atypical:}
& & \hbox{\rm dim}_{N_\up,N_\down} = (8N_\up+4)L-8N_\up^2-8N_\up
\nonu
& &  \qquad \qquad \qquad \qquad \hbox{\rm if}\ \
     N_\down =1,\ 1 \leq N_\up\leq L-1 \ ,
\nonu
& & \hbox{\rm dim}_{N_\up,N_\down} = 4L
\nonu
& & \qquad \qquad \qquad \qquad \hbox{\rm if}\ \
    N_\up=N_\down=0 \ .
\label{dims}
\eea
The numbers $N_\up$ and $N_\down$ are expressed in the
multiplicities $\{M_m,N_1,N_{12},K_m\}$ according to
\bea
N_\up &=& \sum_{m=1}^\infty m(M_m-K_m)-N_{12}
\nonu
N_\down &=& \sum_{m=1}^\infty m(M_m+K_m) - N_1 - N_{12} \ .
\eea

\section{Counting eigenstates}
\setcounter{equation}{0}

With the results of section 2, we are now ready to discuss the
completeness of the Bethe Ansatz solution.
This issue was already addressed in \cite{ek}, where it was
shown that in the thermodynamic limit the leading large-temperature
behavior of the free energy is given by $F/L \sim -T\ln(4)$.
This relation indicates that the number of eigenstates generated
by the Bethe Ansatz is of the order of $4^L$, but due to possible
subleading corrections this is not a rigorous result.

Using (\ref{number}) and (\ref{dims}), we obtain the following
exact expression for the total number of non-vanishing eigenstates
that can be obtained by combining the Bethe Ansatz with the $SU(2|2)$
multiplet structure
\be
\label{totalno}
\# \; (\hbox{\rm eigenstates})=
\sum_{M_m,N_1,N_{12},K_m \geq 0}
   n(M_m,N_1,N_{12},K_m)\, \hbox{\rm dim}_{N_\up,N_\down} \ .
\ee
In this section we will prove that for general $L$ this sum equals
$4^L$. If we then use (as is easily shown) that the eigenstates in
(\ref{totalno}) are all linearly independent, we may conclude that
the Bethe Ansatz does give a complete solution of the model.

As an aside, we remark that the $SU(2|2)$ multiplet structure of
the space of eigenstates of $H_0$ is a consequence of the fact that
we imposed periodic boundary conditions on the model. If we would
instead impose twisted boundary conditions, which in general break
the $SU(2|2)$ symmetry, we would lose the multiplet structure and
obtain a larger set of Bethe Ansatz eigenstates (compare with
\cite{duncan}).

As a first step in the evaluation of (\ref{totalno}), we shall identify
the string multiplicities $\{M_m,N_1,N_{12},K_m\}$ that lead to atypical
representations. It is easily shown that these only arise if all
multiplicities are zero (this gives the vacuum representation of dimension
$4L$) and for the choice $M_1=n$, $N_1=n-1$, with the other multiplicities
vanishing. The contribution of these atypical multiplets to the summation
in (\ref{totalno}) is easily evaluated
\bea
\#({\rm atypical}) &=& 4L + \sum_{n=1}^{L-1}
   \left( \begin{array}{c} L-1 \\ n \end{array} \right)
   \left[ (8n+4)L-8n^2-8n \right]
\nonu
  &=& 2^L \, (L^2-L+2) \ .
\label{atyp}
\eea
Notice that for $L=1,2,3$ this number is actually equal to $4^L$; for
these small lattices {\it all} eigenstates of the hamiltonian belong to
atypical multiplets.

This was the easy part. The hard part is to determine the
contribution from the typical multiplets. It is given by
\bea
&& \#({\rm typical}) =
   \sum^*_{M_n,N_1,N_{12},K_m}
\prod_{m=1}^{\infty} \left( \begin{array}{c}
   \str L - \sum_{n=1}^\infty (t_{nm} M_n)
    + N_1 + N_{12}(2-\delta_{m1}) \\ \str M_m \end{array} \right)
\nonu && \hspace{1.5cm} \times
\left( \begin{array}{c}
   \str \sum_{n=1}^\infty (M_n - K_n) - N_{12} - 1
   \\ \str N_1 \end{array} \right)
\nonu && \hspace{1.5cm} \times
\left( \begin{array}{c}
   \str M_1 + 2 \sum_{n=2}^\infty M_n - N_1 - N_{12} - 2
   \\ \str N_{12} \end{array} \right)
\nonu && \hspace{1.5cm} \times
\prod_{m=1}^\infty \left( \begin{array}{c}
   \str N_1 - \sum_{n=1}^\infty \left( t_{mn} K_n \right)
   \\ \str K_m \end{array} \right)
\nonu && \hspace{1.5cm} \times \,
   16 \left[L+1-2\sum_m (mM_m) + N_1 + 2 N_{12}\right]
   \left[1-2\sum_m(mK_m)+N_1\right] \ ,
\nonu &&
\eea
where the * on the summation symbol indicates that we should
exclude the terms that correspond to atypical multiplets.
Our task is now to evaluate this summation in closed form.

We will tackle this problem in a number of steps, where
we will be using results that were obtained in the analogous
counting problems for simpler models, notably the spin-$\half$
$XXX$ quantum chain.

\vskip 8mm

\noindent \underline{\sc Step 1.}

\vskip 4mm

\noindent
We recall the following identity, which was first written
by Bethe in his original paper on the solution of the
spin-$\half$ $XXX$ quantum chain, and which is valid
for $N\leq [L/2]$
\be
\begin{array}{c}
         \str \\
         \displaystyle{ \sum_{N_1,N_2,\ldots=0}^\infty}
         \\
         \str \scriptstyle{\sum_{m=1}^\infty (m N_m) = N}
         \\
         \str \scriptstyle{\sum_{m=1}^\infty N_m = n}
\end{array}
\prod_{n=1}^\infty
   \left( \begin{array}{c} \Str L-\sum_m (t_{nm} N_m) \\
                           \Str N_n    \end{array}  \right)
= \frac{L-2N+1}{L-N+1}
  \left( \begin{array}{c} L-N+1 \\ n \end{array} \right)
  \left( \begin{array}{c} N-1 \\ n-1 \end{array} \right) .
\label{xxx1}
\ee
Using this identity we can directly perform the summation
over $K_m$, $m=1,2,\ldots$, with the restrictions
$\sum_{m=1}^\infty (mK_m) = N_2$ and $\sum_{m=1}^\infty
K_m = K$. If we shift the integer label of the $M_m$ by
one unit and use the fact that $t_{m+1,n+1}=2+t_{m,n}$,
we can use the same identity (\ref{xxx1}) to perform the
summation over $M_2,M_3,\ldots$ with the restrictions
$\sum_{m=2}^\infty (mM_m)=M$ and $\sum_{m=2}^\infty M_m = q$.
The resulting expression is
\bea
&& \#({\rm typical}) =
   \sum^*_{M_1,M,q,N_1,N_{12},K,N_2}
\left( \begin{array}{c}
   \str L-M_1-2q+N_1+N_{12} \\ \str M_1 \end{array} \right)
\nonu && \hspace{1.5cm} \times
\left( \begin{array}{c}
   \str L-2M_1-M-q+N_1+2N_{12}+1 \\ \str q \end{array} \right)
\left( \begin{array}{c}
   \str M-q-1 \\ \str q-1 \end{array} \right)
\nonu && \hspace{1.5cm} \times
\left( \begin{array}{c}
   \str M_1+q-N_{12}-K-1 \\ \str N_1 \end{array} \right)
\left( \begin{array}{c}
   \str M_1 + 2q - N_1 - N_{12} - 2 \\ \str N_{12} \end{array} \right)
\nonu && \hspace{1.5cm} \times
\left( \begin{array}{c}
   \str N_1 - N_2 + 1\\ \str K \end{array} \right)
\left( \begin{array}{c}
   \str N_2 - 1 \\ \str K-1 \end{array} \right)
\nonu && \hspace{1.5cm} \times
   16 \, \frac{(L-2M_1-2M+N_1+2 N_{12}+1)^2}
           {L-2M_1-M-q+N_1+2N_{12}+1}
   \, \frac{(N_1-2N_2+1)^2}{N_1-N_2+1}  \ .
\eea

\vskip 8mm

\noindent \underline{\sc Step 2.}

\vskip 4mm

\noindent
The summation over $N_2$ can be performed as follows
\bea \lefteqn{
\sum_{N_2=0}^{[N_1/2]}
\left( \begin{array}{c} \str N_1-N_2+1 \\ \str K \end{array} \right)
\left( \begin{array}{c} \str N_2-1 \\ \str K-1 \end{array} \right)
\frac{(N_1-2N_2+1)^2}{N_1-N_2+1}}
\nonu &=&
\sum_{N_2=0}^{[N_1/2]} (N_1-2N_2+1) \left[
\left( \begin{array}{c} \str N_1-N_2+1 \\ \str K \end{array} \right)
\left( \begin{array}{c} \str N_2-1 \\ \str K-1 \end{array} \right)
-
\left( \begin{array}{c} \str N_1-N_2 \\ \str K-1 \end{array} \right)
\left( \begin{array}{c} \str N_2 \\ \str K \end{array} \right)
\right]
\nonu &=& - \sum_{N_2=0}^{N_1+1} (N_1-2N_2+1)
\left( \begin{array}{c} \str N_1-N_2 \\ \str K-1 \end{array} \right)
\left( \begin{array}{c} \str N_2 \\ \str K \end{array} \right)
\nonu &=&
\left( \begin{array}{c} \str N_1+1 \\ \str 2K+1 \end{array} \right) ,
\eea
where we used the relations
\bea &&
\sum_{N_2=0}^{N_1+1}
\left( \begin{array}{c} \str N_1-N_2 \\ \str K-1 \end{array} \right)
\left( \begin{array}{c} \str N_2 \\ \str K \end{array} \right)
=
\left( \begin{array}{c} \str N_1+1 \\ \str 2K \end{array} \right)
\nonu &&
\sum_{N_2=0}^{N_1+1} (N_2-K)
\left( \begin{array}{c} \str N_1-N_2 \\ \str K-1 \end{array} \right)
\left( \begin{array}{c} \str N_2 \\ \str K \end{array} \right)
= (K+1)
\left( \begin{array}{c} \str N_1+1 \\ \str 2K+1 \end{array} \right) ,
\eea
which are easily proved. The sum over $K$ is performed as
\be
\sum_K
\left( \begin{array}{c} \str N_1+1 \\ \str 2K+1 \end{array} \right)
\left(\begin{array}{c} \str X-K \\ \str N_1 \end{array} \right)
= \left( \begin{array}{c} \str 2X-N_1+1 \\ \str N_1 \end{array} \right)
\ ,
\ee
with $X=M_1+q-N_{12}-1$. We then change variables from
$\{M_1,M,N_1,q,N_{12}\}$ to $\{A,B,C,q,N_{12}\}$, where
\be
A = 2M_1+2q-N_1-2N_{12}\ , \qquad B=M-q\ ,
\qquad C = M_1+q-N_1-N_{12}-1 \ .
\ee
With that, the summation over $N_{12}$ becomes elementary
and we obtain
\bea
&& \#({\rm typical}) =
   \sum^*_{A,B,C,q}
\left( \begin{array}{c}
   \str L-A-B+1 \\ \str q \end{array} \right)
\left( \begin{array}{c}
   \str B-1 \\ \str q-1 \end{array} \right)
\nonu && \hspace{1.5cm} \times
\left( \begin{array}{c}
   \str L-2 \\ \str A-2 \end{array} \right)
\left( \begin{array}{c}
   \str A-1 \\ \str 2C+1 \end{array} \right)
   \, 16 \, \frac{(L-A-2B+1)^2}{L-A-B+1} \ .
\eea

\vskip 8mm

\noindent \underline{\sc Step 3.}

\vskip 4mm

\noindent
Before we continue, let us specify the range of the
variables $q$ and $C$. The summand in the above is nonvanishing
when $q\geq 0$ and $C\geq 0$. However, we should exclude the
terms that give rise to atypical representations of $SU(2|2)$;
it can easily be seen that those correspond precisely to
the choice $q=C=0$. The summations should thus be carried out
over the regions $q\geq 1$, $C\geq 0$ and $q=0$, $C\geq 1$.
Performing the summation over $C$ we obtain
\bea
&& \#({\rm typical}) =
   \sum_{A,B,q}
\left( \begin{array}{c}
   \str L-A-B+1 \\ \str q \end{array} \right)
\left( \begin{array}{c}
   \str B-1 \\ \str q-1 \end{array} \right)
\left( \begin{array}{c}
   \str L-2 \\ \str A-2 \end{array} \right)
\nonu && \hspace{1.5cm} \times
   16 \, \frac{(L-A-2B+1)^2}{L-A-B+1}
   \left\{ \begin{array}{cc} \Str 2^{A-2} & \hbox{\rm if}\ \ q\geq 1 \\
                            \Str 2^{A-2}-(A-1) & \hbox{\rm if}\ \ q=0 \ .
          \end{array} \right.
\eea
The summation over $q$ then gives
\be
\#({\rm typical}) = I - II \,
\ee
where
\bea
I &=& \sum_{A,B} 16 \, \frac{(L-A-2B+1)^2}{L-A-B+1} \,
   \left( \begin{array}{c} \str L-A \\ \str B \end{array} \right)
   \left( \begin{array}{c} \str L-2 \\ \str A-2 \end{array} \right)
   2^{A-2}
\nonu &=&
\sum_{A,B} 16 \, (L-A-2B+1) \, \left[
   \left( \begin{array}{c} \str L-A \\ \str B \end{array} \right) -
   \left( \begin{array}{c} \str L-A \\ \str B-1 \end{array} \right)
   \right]
   \left( \begin{array}{c} \str L-2 \\ \str A-2 \end{array} \right)
   2^{A-2}
\nonu &&
\eea
and
\be
II = \sum_A 16 \, (L-A+1) \, (A-1)
   \left( \begin{array}{c} \str L-2 \\ \str A-2 \end{array} \right) \ .
\ee
We can now use the following lemma, which originates from
the counting of states in the spin-$\half$ $XXX$-model
\be
  \sum_{M=0}^{[N/2]}
  \left[ \left( \begin{array}{c} N \\ M \end{array} \right)
         - \left( \begin{array}{c} N \\ M-1 \end{array} \right)
  \right] (N - 2 \, M +1) = 2^N \, ,
\label{xxx2}
\ee
to obtain
\be
I = \sum_A 16 \,
   \left( \begin{array}{c} \str L-2 \\ \str A-2 \end{array} \right)
   \, 2^{L-A} \, 2^{A-2} = 2^{2L} = 4^L  \ .
\ee
The second term can be evaluated as
\bea
II &=& \sum_A 16 \, (L-A+1) \, (A-1)
   \left( \begin{array}{c} \str L-2 \\ \str A-2 \end{array} \right)
\nonu
   &=& 2^L \, (L^2-L+2)
\eea
and we find
\be
\#({\rm typical}) = 4^L - 2^L(L^2-L+2) \ .
\label{typres}
\ee
Comparing with (\ref{atyp}), we see that the contribution
from the atypical multiplets precisely cancels the second term
in (\ref{typres}) so that the full result of the summation
(\ref{totalno}) is given by
\be
\#(\hbox{\rm eigenstates}) = 4^L  \ .
\ee
This is the desired result.

\vskip 4mm

\noindent {\it Acknowledgements.}\ It is a great pleasure to
thank Vladimir Korepin and Fabian E\char'31 ler for collaboration
on this model and for illuminating discussions on the subject of this
paper. This research was supported by DOE grant DE-AC02-76ER-03072.

\end{document}